# پردازش و تحلیل کامپیوتری تصاویر سی‌تی‌اسکن ریه در راستای ارایه‌ی معیاری جهت تشخیص بیماری انسدادی ریوی

**محمدپارسا حسینی[۱]، دکتر حمید سلطانیان‌زاده[۲]، دکتر شهرام اخلاقپور[۳]**

## چکیده

**مقدمه:** بیماری انسدادی ریه یکی از شایع‌ترین و خطرناک‌ترین بیماری‌های ریوی است. پیش‌بینی می‌شود طی سال‌های آتی، این بیماری به عنوان سومین بیماری کشنده‌ی دنیا شناخته شود. ارایه‌ی روش‌های جدید در راستای تشخیص بیماری کمک ارزنده‌ای به پزشکان و بیماران خواهد کرد.

**روش‌ها:** ده بیمار مبتلا به بیماری انسدادی ریه (۶ مرد و ۴ زن با میانگین سنی ۴۹/۸ سال) بر طبق معاینات کلینیکی و نتایج تست تنفسی وارد مطالعه گردیدند. در همین راستا، ده شخص سالم (۶ مرد و ۴ زن با میانگین سنی ۴۵/۴ سال) نیز به عنوان گروه شاهد مورد بررسی قرار گرفتند. تصاویر سی‌تی‌اسکن ریه‌ی این دو گروه، توسط کامپیوتر پردازش و تحلیل گردید.

**یافته‌ها:** میزان ارتجاع پذیری و الاستیسیته‌ی نسج ریه‌ی نمونه‌ها از طریق تحلیل کامپیوتری تصاویر بررسی شد. میانگین نرمالیزه شده‌ی این ویژگی در گروه بیمار ۲۱/۶ درصد و در گروه شاهد ۴۰/۷ درصد به دست آمد. همچنین میانگین نرمالیزه شده‌ی تغییرات دوز جذبی در پنجره‌های مربعی شکل ۱۰ پیکسلی در تصاویر بازدمی به عنوان معیاری جهت بررسی میزان گیرافتادگی هوا محاسبه گردید. متمایز بودن این معیارها توسط آزمون آماری t-student با ($P < ۰/۰۵$) اثبات گردید.

**نتیجه‌گیری:** این مطالعه نشان داد که ارتجاع پذیری و دوز جذبی نسج ریه که توسط پردازش و تحلیل کامپیوتری تصاویر سی‌تی‌اسکن به دست آمد، می‌تواند به عنوان یک معیار در راستای تشخیص و بیان شدت بیماری به کار گرفته شود. از این‌رو، می‌توان از این روش پیشنهادی در کنار سایر معیارها برای تشخیص بیماری انسدادی ریه استفاده نمود.

## مقدمه

بیماری مزمن انسدادی ریه (COPD یا Chronic obstructive pulmonary disease) یکی از شایع‌ترین بیماری‌های ریوی است که در بزرگسالان ایجاد می‌شود. برونشیت مزمن، آمفیزم و آسم اقسام این بیماری هستند. در آمفیزم، بزرگی برگشت ناپذیری در فضای هوایی در محل اتصال برونش‌ها همراه با از بین رفتن راه‌های هوایی آن نقاط ایجاد می‌گردد (۳-۱). در برونشیت با ایجاد التهاب در برونش‌ها، عملکرد ارگان کاهش می‌یابد و در حالت شدید آن یعنی در برونکستازی اتساع برگشت ناپذیر محلی در درخت برونشی همراه با بی‌نظمی‌های گسترده ی ناشی از التهاب شدید ایجاد می‌شود (۶-۴). در آسم، مجاری هوایی در بیمار به صورت غیر طبیعی و به طور متناوب تنگ می‌شود. انسداد ایجاد شده در این بیماری به دنبال التهاب مسیر هوایی و افزایش موکوس در راه‌های

[۱] کارشناس ارشد، گروه مهندسی برق و کامپیوتر، واحد علوم و تحقیقات، دانشگاه آزاد اسلامی، تهران، ایران
[۲] پژوهشگر ارشد، آزمایشگاه تحلیل تصاویر، بخش رادیولوژی، بیمارستان هنری فورد، دیترویت، آمریکا و استاد تمام، قطب علمی کنترل و پردازش هوشمند، دانشکده‌ی مهندسی برق و کامپیوتر، دانشگاه تهران، تهران، ایران
[۳] دانشیار، گروه رادیولوژی، بیمارستان سینا، دانشگاه علوم پزشکی تهران، تهران، ایران
**نویسنده‌ی مسؤول:** محمدپارسا حسینی
Email: mp.hosseini@srbiau.ac.ir

هوایی ریه‌ها رخ می‌دهد، که سبب کاهش جریان هوا خواهد شد. شکل ۱ تغییرات راه‌های هوایی در برونشیت و تغییرات فضای هوایی در آمفیزم را نشان می‌دهد. بیماری فوق چهارمین علت مرگ و میر در آمریکا و از دلایل عمده‌ی مرگ و میر در کشورهای در حال توسعه است. بر طبق پیش‌بینی‌های انجام شده، بیماری انسدادی ریه سومین علت مرگ و میر آینده خواهد بود (۸-۷). آزمایش‌هایی همچون اسپیرومتری، گرفتن نمونه‌ی گاز خون شریانی و عکس‌برداری از قفسه‌ی سینه جهت تشخیص این بیماری انجام می‌پذیرد که در نهایت پزشک باید با بررسی نتایج و استفاده از تجارب خویش، تشخیص خود را ارایه نماید. عواملی همچون خطای انسانی و عدم دسته‌بندی صحیح بر حسب شدت بیماری، نیاز به یک روش مکانیزه برای تشخیص بیماری و کلاسه‌بندی بیماران را مبرم و کارساز می‌سازد.

از میان روش‌های مختلف تصویربرداری پزشکی، روش سی‌تی‌اسکن جهت تصویربرداری از ریه‌ها مرسوم می‌باشد (۱۳-۱۱). در سال ۲۰۰۹ میلادی Van Beek و Hoffman به بررسی روش‌های مختلف برای افزایش کنتراست و اخذ جزئیات بیشتر ریه‌ها در تصویربرداری از بیماران انسداد ریوی پرداختند (۸). حسینی و همکاران به تحلیل تصاویر سی‌تی‌اسکن ریه پرداخته‌اند (۱۰-۹). در این پژوهش‌ها تغییرات حجمی تصاویر جهت شناسایی نشانه‌های بیماری مورد توجه بوده است.

شناسایی بیماری فوق و بیان درصد شدت آن، مرحله‌ی مهمی در تشخیص و معالجه‌ی بیمار خواهد بود. در این پژوهش، روشی نوین در تشخیص این نوع بیماری‌ها با استفاده از تصاویر گرفته شده از بیمار مطرح می‌گردد که علاوه بر تشخیص و تعیین احتمال وجود بیماری در شخص، قابلیت بیان خودکار شدت بیماری را نیز خواهد داشت.

## روش‌ها

پس از تأیید طرح توسط معاونت پژوهشی واحد علوم و تحقیقات و کمیته‌ی اخلاق پزشکی مرکز تصویربرداری پزشکی نور، مطالعه‌ی کارآزمایی بالینی تصادفی طی زمستان ۱۳۸۸ و بهار ۱۳۸۹ در مرکز تصویربرداری پزشکی نور واقع در خیابان مطهری تهران، انجام پذیرفت. بدین منظور، پایگاه داده‌ای ۲۰ نفری شامل ۱۰ فرد سالم و ۱۰ بیمار تشکیل گردید.

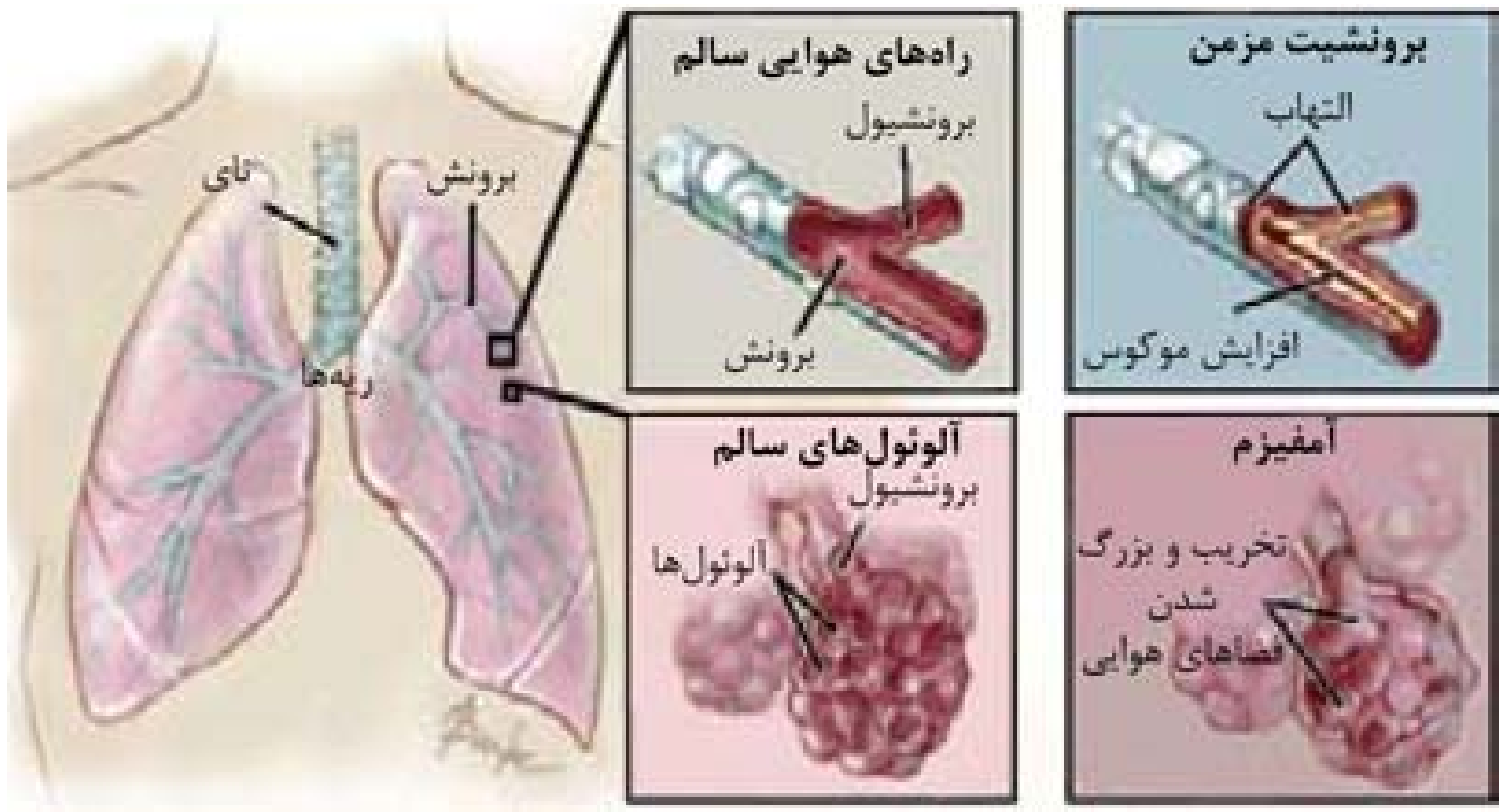

**شکل ۱. تصویر راه‌های هوایی و آلوئول‌های سالم و آسیب دیده در بیماری‌های انسدادی**

تصاویر مورد استفاده توسط دستگاه سی‌تی‌اسکن زیمنس ۶۴ برشه با قدرت تفکیک فضایی بالا (HRCT یا High resolation CT) و ضخامت ۱ میلی‌متری، تهیه گردید. در شکل ۲ نمونه‌ای از تصاویر دمی و بازدمی مورد مطالعه نمایش داده شده است.

با توجه به ساختار پیچیده و متغیر ریه در طول و عرض روش‌های متعدد بخش‌بندی مورد ارزیابی قرار گرفت و در نهایت روش بخش‌بندی کانتورهای فعال (Active countors) که جزء روش‌های ساختاری بخش‌بندی تصاویر محسوب می‌شوند، انتخاب گردید. در این روش، انرژی کانتور از سه انرژی درونی، خارجی و محدودیت تشکیل شده است که به مجموع انرژی‌های فوق، انرژی مار (Snake) می‌گویند. با توجه به شکل، شدت روشنایی و دقت مطلوب، انرژی فوق تعریف می‌شود. هنگامی که به لبه‌های شکل برسیم، انرژی فوق کمینه می‌گردد؛ این لبه‌ها، نقطه‌ی توقف عملیات بخش‌بندی خواهد بود.

$$V = [v_1, \dots, v_n] \qquad (۱)$$

$$v_i \in I = \{(x, y) | x, y = 1,2,\dots,m\} \qquad (۲)$$

$$E_{snake} = \int_0^1 E_{snake}(v(s))ds$$
$$= \int_0^1 E_{Internal}(v(s))ds + \int_0^1 E_{External}(v(s))ds + \int_0^1 E_{Constra\,int}(v(s))ds \qquad (۳)$$

$V$ بیان‌گر نقاط کنترلی یا همان نقاط اولیه ی شروع عملیات است. همانطور که در معادله‌ی ۲ اشاره شده است، نقاط کنترلی باید با توجه به اندازه ی تصویر که در این‌جا m در m فرض شده است، از پیکسل‌های تصویر انتخاب شوند. در معادله‌ی ۳ به انرژی‌های تشکیل دهنده ی انرژی مار اشاره شده است. در روش فوق با تکرارهای متناوب، کانتور از یک نقطه‌ی کنترلی شروع به رشد می‌نماید تا جایی که انرژی کانتور کمینه گردد (۱۴). بدین صورت لبه‌های نسج ریه استخراج خواهد گردید.

در مرحله ی بعد باید تصاویر متناظر در هر بیمار به دست بیاید که این امر با توجه به شکل ریه و ساختارهای آناتومیکی آن انجام می‌شود. هر ریه تا حدودی مخروطی شکل می‌باشد و قاعده ی هر ریه مقعر است و روی دیافراگم قرار دارد. کناره‌ی قدامی ریه نازک است و قلب را می‌پوشاند و کناره‌ی قدامی ریه ی چپ در ناحیه ی قلب دارای یک بریدگی به نام بریدگی قلبی می‌باشد. کناره‌ی خلفی ریه ضخیم است و در کنار ستون فقرات قرار دارد که دارای یک سطح دنده‌ای محدب متناسب با جدار توراکس می‌باشد. با معلومات فوق تناظر تصاویر به دست آمد.

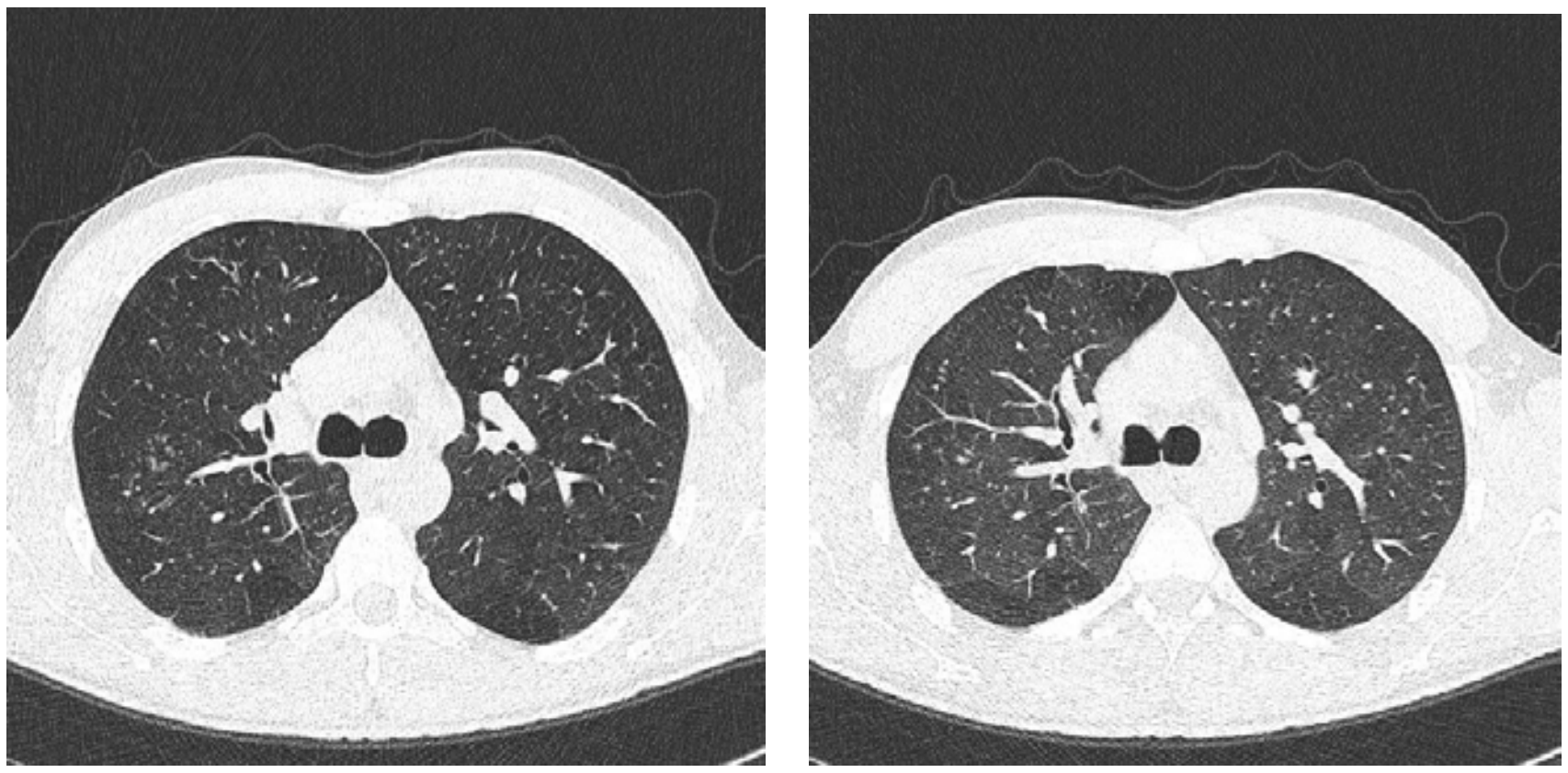

**شکل ۲. نمونه‌ای از تصاویر اولیه‌ی مورد استفاده. راست: بازدم، چپ: دم (مرد، ۲۷ ساله، گروه شاهد)**

در مرحله‌ی بعد تصاویر پیش پردازش شده جهت استخراج ویژگی، مورد ارزیابی قرار گرفتند. بدین منظور تصاویر دو گروه بیمار و شاهد بررسی شد تا ویژگی‌هایی متناسب با بیماری فوق استخراج گردد. با توجه به تحلیل‌های انجام شده و میزان اثربخشی نتایج، ویژگی تغییرات سطح و حجم و میزان تغییرات دوز جذبی، به عنوان ویژگی تشخیص بیماری و سلامت در شخص معرفی شدند. این ویژگی‌ها به طور مستقیم نشان دهنده‌ی تغییرات الاستیسیته و کش‌سانی نسج ریه و میزان گیرافتادگی هوا (Air-trapping) می‌باشند. برای حذف تأثیر جثه‌ی افراد در نتایج هر یک از ویژگی‌ها با مقدار دمی خود نرمالیزه گردیدند. بدین صورت، پارامتری بین ۰ و ۱ به دست خواهد آمد که هر چه مقدار فوق نزدیک‌تر به صفر باشد، بیان‌گر میزان انسداد بیشتری در آن موضع خواهد بود. شکل ۳ نتایج میزان تغییرات را در دو حالت نشان می‌دهد.

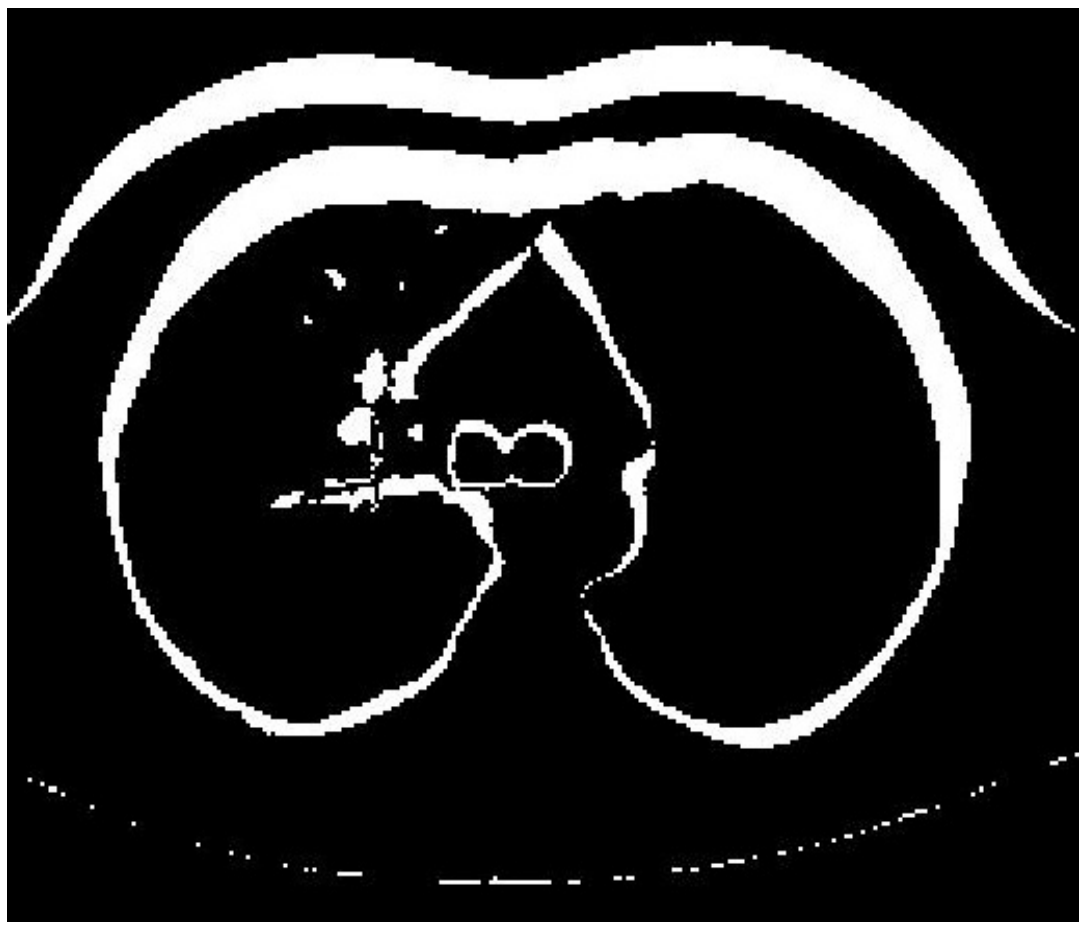

**شکل ۳. مقایسه‌ی سطحی تصاویر دمی و بازدمی**

جهت شناسایی میزان گیرافتادگی هوا، نسج ریه در بازدم به پنجره‌های مربعی شکل که در هر جهت ۱۰ پیکسل کشیدگی دارند، تقسیم گردید. با بررسی مقادیر میانگین دوز جذبی در این پنجره‌ها، تغییرات مقدار Hounsfield به دست آمد و با بیشترین مقدار، نرمالیزه گردید. در شکل ۴ نمونه‌ای از این پنجره‌ها نمایش داده شده است.

در نهایت به منظور کلاس‌بندی تصاویر، توزیع آماری این ویژگی‌ها به دست آمد و آستانه‌ی سخت تفکیک کننده‌ی دو کلاس برای هر ویژگی بیان گردید. به عبارت دیگر، طبقه‌بندی کننده با استفاده از این دادگان آموزش یافت و سپس دادگان جدید با مقایسه با آستانه‌ی به دست آمده، به دو گروه سالم و بیمار طبقه‌بندی شدند.

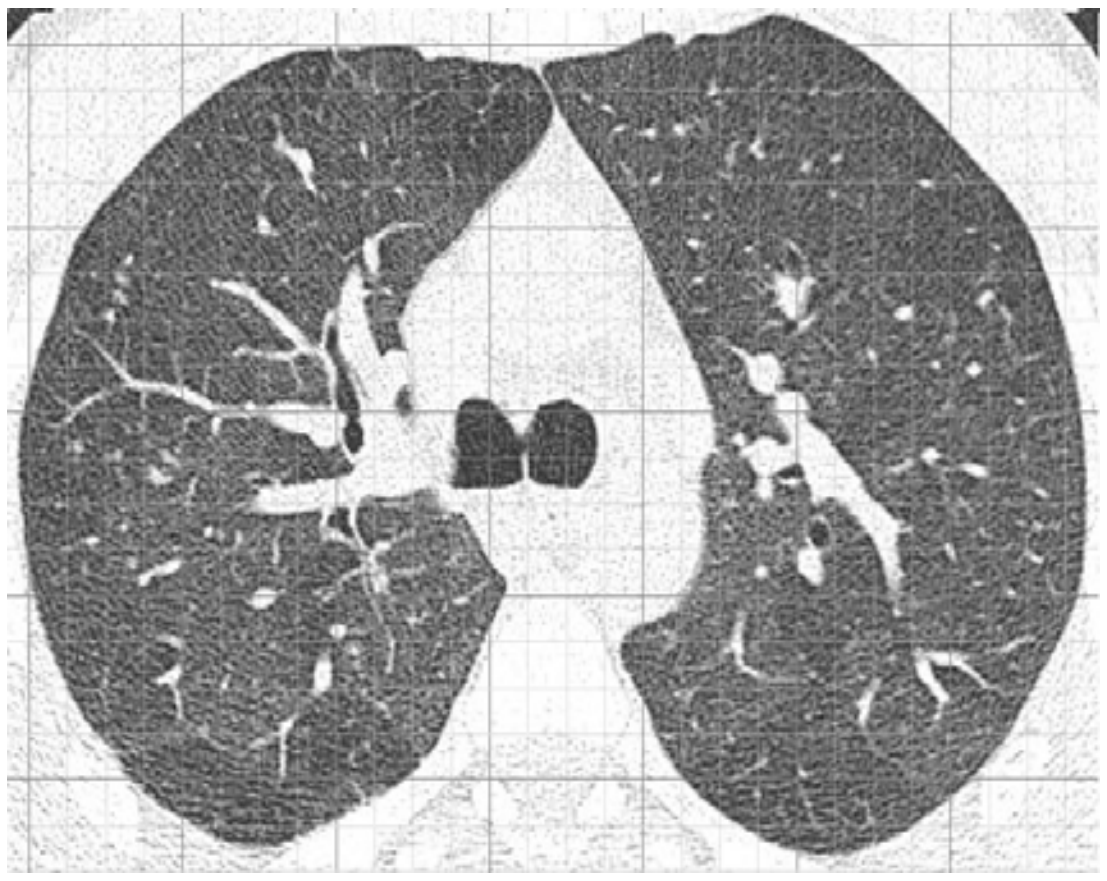

**شکل ۴. نمونه‌ای از پنجره‌های مورد استفاده در محاسبه‌ی دوز جذبی معادل (جهت وضوح شکل، پنجره‌ها بزرگ‌تر از اندازه‌ی واقعی نمایش داده شده‌اند.)**

## یافته‌ها

نتایج حاصل از استخراج ویژگی‌ها بر روی تصاویر گروه شاهد و بیمار بررسی گردیدند. جدول ۱ مقدار ویژگی‌های پیش‌گفت بر روی تصاویر نمونه‌های این تحقیق را نشان می‌دهد. به منظور بررسی شرایط یکسان برای تغییرات سطحی، تصاویر مقاطع کارینا استفاده گردیدند.

**جدول ۱. دو ویژگی استخراج شده از تصاویر افراد بیمار و سالم**

| ویژگی دوم | ویژگی اول | تشخیص پزشک | جنس زن | جنس مرد | سن | شماره‌ی نمونه |
|---|---|---|---|---|---|---|
| ۰/۰۳۵ | ۰/۰۹۷ | بیمار | | * | ۴۰ | ۱ |
| ۰/۱۱۸ | ۰/۰۷۳ | بیمار | | * | ۳۱ | ۲ |
| ۰/۴۳۷ | ۰/۴۲۷ | بیمار | * | | ۳۸ | ۳ |
| ۰/۲۵۱ | ۰/۲۸۶ | بیمار | | * | ۲۱ | ۴ |
| ۰/۲۲۷ | ۰/۱۸۱ | بیمار | * | | ۴۲ | ۵ |
| ۰/۲۷۶ | ۰/۲۷۴ | بیمار | | * | ۷۴ | ۶ |
| ۰/۲۵۷ | ۰/۲۴۱ | بیمار | | * | ۴۸ | ۷ |
| ۰/۰۶۸ | ۰/۰۴۳ | بیمار | * | | ۶۷ | ۸ |
| ۰/۰۵۵ | ۰/۰۰۹ | بیمار | * | | ۸۸ | ۹ |
| ۰/۴۳۳ | ۰/۳۹۹ | بیمار | | * | ۴۹ | ۱۰ |
| ۰/۴۷۶ | ۰/۴۱۸ | سالم | * | | ۴۳ | ۱۱ |
| ۰/۴۵۸ | ۰/۳۹۲ | سالم | * | | ۲۲ | ۱۲ |
| ۰/۲۶۷ | ۰/۲۳۹ | سالم | * | | ۴۷ | ۱۳ |
| ۰/۳۲۸ | ۰/۲۴۷ | سالم | | * | ۴۴ | ۱۴ |
| ۰/۴۶۸ | ۰/۴۸۲ | سالم | * | | ۴۳ | ۱۵ |
| ۰/۴۹۵ | ۰/۴۷۶ | سالم | | * | ۶۵ | ۱۶ |
| ۰/۳۹۱ | ۰/۳۷۸ | سالم | | * | ۴۸ | ۱۷ |
| ۰/۴۸۵ | ۰/۳۷۷ | سالم | | * | ۵۳ | ۱۸ |
| ۰/۳۶۳ | ۰/۳۲۵ | سالم | | * | ۴۵ | ۱۹ |
| ۰/۳۳۵ | ۰/۳۱۳ | سالم | | * | ۴۴ | ۲۰ |

در مورد ویژگی اول در جامعه‌ی آماری بررسی شده، در کلاس بیمار میانگین ۰/۲۰۳ با انحراف معیار ۰/۱۵۵ و در کلاس سالم میانگین ۰/۳۶۵ با انحراف معیار ۰/۰۸۶ به دست آمد. با استفاده از تئوری Bayes و تخمین پارامتریک تابع چگالی احتمال به صورت تابع گوسی، سطح آستانه‌ی سخت تشخیص بیماری و سلامت ۰/۲۶۵ به دست آمد. برای ویژگی دوم در کلاس بیمار میانگین ۰/۲۱۶ با انحراف معیار ۰/۱۵۳ و در کلاس سالم میانگین ۰/۴۰۷ با انحراف معیار ۰/۰۸۰ به دست آمد. استفاده از قانون Bayes در این ویژگی سطح آستانه‌ی تشخیصی را ۰/۳۰۵ نشان داد. هر چقدر مقدار ویژگی به دست آمده بزرگ‌تر از آستانه باشد، بیانگر سلامت و احتمال کمتر وجود بیماری در شخص است.

در انتها آزمون آماری t-student بر روی دادگان اعمال گردید. برای ویژگی اول درجه‌ی آزادی توزیع آماری دادگان ۱۴/۰۶ بود. با تعریف سطح ریسک ۰/۰۵ فرض خلف یکسان بودن میانگین‌های دو کلاس با خطای ۰/۰۰۶ رد می‌شود. به همین ترتیب، برای ویژگی دوم درجه‌ی آزادی توزیع ۱۳/۵۴ به دست آمد که بیانگر استقلال میانگین‌های دو کلاس با درصد خطای ۰/۰۰۱۹ خواهد بود.

## بحث

در این پژوهش با توجه به نیازی که در زمینه‌ی تشخیص بیماری‌های انسدادی برای پزشکان و بیماران وجود داشت، تصاویر سی‌تی‌اسکن ریه ی بیماران انسدادی و افراد سالم پردازش و تحلیل شد و روشی نوین برای شناسایی و تقسیم‌بندی خودکار بیماران پیشنهاد گردید. به منظور فوق، ابتدا روش‌های مختلف بخش‌بندی تصاویر پزشکی مورد ارزیابی قرار گرفت و در نهایت، روش کانتورهای فعال با توجه به بخش‌بندی قابل قبول نسج ریه انتخاب گردید. سپس تصاویر دمی و بازدمی با توجه به ساختارهای آناتومیکی حجمی ریه متناظر شدند و تغییرات سطح و حجم ریه و همچنین میزان تغییرات دوز جذبی بافت ریه به صورت نرمالیزه بیان گردید. در انتها، نتایج کار با دادگان اخذ شده مورد ارزیابی قرار گرفت و سطح آستانه‌ی تفکیک سالم و بیمار به دست آمد. با بررسی تغییرات ویژگی‌ها بر روی تصاویر بیماران و افراد سالم، کارایی روش پیشنهادی در اکثر موارد به خوبی مشخص می‌باشد. از قابلیت‌های این روش عدم نیاز به حضور پزشک متخصص جهت شناسایی بیماری می‌باشد. بنابراین در مقایسه با کارهای مشابه در این

زمینه از کارایی بالاتری برخوردار خواهد بود.

## تشکر و قدردانی

به این وسیله از همکاری کلینیک تصویربرداری نور، انستیتو پرتوپزشکی نوین و قطب علمی کنترل و پردازش هوشمند دانشگاه تهران که ما را در انجام این پژوهش یاری نمودند، تشکر و قدردانی می‌نماییم.

# Computerized Processing and Analysis of CT Images for Developing a New Criterion in COPD Diagnosis

Mohammad Parsa Hosseini MSc[1], Hamid Soltanian Zadeh PhD[2], Shahram Akhlaghpoor MD[3]

## Abstract

**Background:** Chronic obstructive pulmonary disease (COPD) is one of the most prevalent and dangerous pulmonary diseases in the world. It is forecasted that COPD will be the third deadly disease in the future. Therefore, developing non-invasive methods for diagnosis of the disease would be helpful for physicians and patients.

**Methods:** Based on clinical investigations and spirometry tests, ten adult patients with COPD (6 male and 4 female) with mean age of 49.8 years were enrolled as the case group. In addition, ten age and sex-matched healthy, non-COPD individuals (6 male and 4 female) with mean age of 45.4 years were recruited as the controls. Lung CT-scan images of the subjects were processed and analyzed by a computer to find a relationship.

**Findings:** The elasticity of lung parenchyma variation was obtained with digital image processing. The normalized average of this pattern was found to be 21.6% in patients and 40.7% in controls. In addition, normalized mean value of Hounsfield unit variations in square 10 pixel × 10 pixel windows in the expiratory images were calculated as a parameter of air-trapping in COPD. Differences between the groups were shown by student t-test ($P < 0.05$).

**Conclusion:** This study showed that the variation of lung parenchyma elasticity and Hounsfield units are found by processing and analysis of the full inspiration and expiration images. These factors can be used as criteria in diagnosis of COPD. Moreover, the severity of the disease can be presented by the proposed method.

**Keywords:** Air-trapping, Chronic obstructive pulmonary disease, Image processing and analysis, CT-scan lung images

[1] Department of Electrical and Computer Engineering, Science and Research Branch, Islamic Azad University, Tehran, Iran
[2] Image Analysis Laboratory, Department of Radiology, Henry Ford Health System, Detroit, Michigan, USA AND Professor, Control and Intelligent Processing Center of Excellence (CIPCE), School of Electrical and Computer Engineering, Tehran University, Tehran, Iran
[3] Associate Professor, Department of Radiology, Sina Hospital, Tehran University of Medical Sciences, Tehran, Iran
**Corresponding Author:** Mohammad Parsa Hosseini MSc, Email: mp.hosseini@srbiau.ac.ir